# Crystal structure and epitaxy of $Bi_2Te_3$ films grown on Si

Jihwey Park, Yeong-Ah Soh, and G. Aeppli

London Centre for Nanotechnology, University College London, London WC1H 0AH, United Kingdom

S. R. Bland

Department of Physics, Durham University, Durham, DH1 3LE, United Kingdom

Xie-Gang Zhu, Xi Chen, Qi-Kun Xue, and Francois Grey

Department of Physics, Tsinghua University, Beijing 100084, P. R. China

**ABSTRACT**

We report comprehensive x-ray diffraction studies of the crystal structure and epitaxy of thin films of the topological insulator $Bi_2Te_3$ grown on Si (1 1 1). The films are single crystals of high crystalline quality, which strongly depends on that of their substrates, with in-plane epitaxial relationships of $Bi_2Te_3$ [2 1 -3 0] || Si [1 -1 0] and $Bi_2Te_3$ [0 1 -1 0] || Si [1 1 -2] along which the lattices of 1×3 $Bi_2Te_3$ and 2×2 Si supercells are well matched. As the samples age, we observe loss of crystalline $Bi_2Te_3$ film thickness accompanied with roughening of the crystalline interfaces, formation of new crystalline phases as well as compositional and structural modification of the Si substrate, consistent with the diffusion of Te into the Si substrate.

Recently, new quantum materials called topological insulators, in which metallic surface states located in the band gap of their insulating bulk are protected by time-reversal symmetry, have been discovered[1-4] and attracted considerable attention due to fascinating properties such as almost dissipationless surface transport and potential for "spintronic" applications.[5,6] $Bi_2Te_3$ (BT) is such a topological insulator with the advantages of a nominally stoichiometric material and a large bulk gap.[3] Bulk single crystals of this substance, however, have a considerable disadvantage over the MBE grown II-VI heterostructures for which many of the interesting effects have been first observed[1] in that transport, which is supposed to occur only through the surface states, also involves unwanted bulk channels due to crystal imperfections.[3,7] For electronic device research and applications it is required not only to eliminate these charge carriers but also to fabricate low-dimensional nanostructures. Therefore, growth of undoped thin films of topological insulators with low defect concentration, in which the Fermi level intersects only the metallic surface states, is essential and has been attempted using various methods. Large progress has been made in thin film growth of topological insulators on Si (1 1 1) substrates using the molecular beam epitaxy (MBE) technique.[8] While the authors of ref. 8 have demonstrated the crystal structure of the surface to correspond to that of BT using RHEED, and shown the presence of metallic surface states using ARPES, so far, no quantitative structural characterization of these films has been performed. Also, the effect of film growth on the substrates, which could even result in doping-induced substrate metallicity, is largely unknown.

In this letter we report the crystal structures of BT thin films grown on Si (1 1 1) substrates characterized by x-ray diffraction (XRD). The XRD results show that MBE yields epitaxial films with good crystalline quality, which is strongly dependent on that of the Si substrate. However, we have found that the film boundaries undergo ageing and also that new phases form with time. We observe strong indications of oxidation of the film surface and diffusion of Te into the Si substrate, which may explain the ageing and formation of new phases, and this implies that electrical transport and infrared measurements on thin films will be strongly affected by the presence of oxide layers, Te vacancies in the films, and Te dopants in silicon.

In the epitaxial growth of thin films the lattice mismatch between two adjacent materials is an important factor influencing the quality of the films as established by metrics such as the electronic mean free path. Single crystal BT has a hexagonal structure in which the stacking sequence of the layer along the [0 0 0 1] direction is Te-Bi-Te-Bi-Te as shown in Fig. 1(a), which forms a quintuple layer (QL) and allows the layer-by-layer growth of BT thin films by MBE.[8] The coupling between two QLs is much weaker than that between two atomic layers within a QL.[9] Both Si (1 1 1) and BT (0 0 0 1) planes exhibit a hexagonal lattice. However, the distances between nearest atoms in the Si (1 1 1) and BT (0 0 0 1) planes are 3.84 Å and 4.38 Å, respectively. Although upon initial inspection, this large lattice mismatch of 14% can lead to the conclusion that the crystal quality will be poor, we show that the two materials are very well lattice matched and yield BT films of high crystalline quality.

The MBE growth of our samples was carried out in an Omicron MBE chamber. The substrates we used were clean Si (1 1 1)-7×7, which could be attained by annealing the substrate at 450 °C for 5 hours and subsequent cyclical heating to 1250 °C.[10,11] The room temperature resistivity of the Si substrates used for growth of sample A1 and A2 is 1 kΩ·cm and 0.02 kΩ·cm for samples B1 and B2 (B-doped Si). Standard Knudsen diffusion cells were used to evaporate Bi (99.9999 %) and Te (99.999 %) sources. All of our thin film samples were prepared at a substrate temperature of 270 °C and a $Te_2$/Bi flux ratio of larger than 8 with a growth rate of 2 QL per minute. Samples A1 and A2 were grown with nominal thicknesses of 50 nm and 150 nm in June 2010 and samples B1 and B2 were grown with nominal thicknesses of 50 nm and 130 nm in April 2010, respectively. All the samples were stored at ambient atmosphere. The XRD experiments were performed twice, in August 2010 and in April 2012, at the X20A beamline of the National Synchrotron Light Source at Brookhaven National Laboratory. A double Ge (1 1 1) crystal was used as a monochromator for x-rays at 8.06 keV and a Si (1 1 1) analyzer was in front of a scintillation detector. All measurements were conducted at room temperature.

Figures 1(b)-(e) show the XRD patterns measured on sample B2 in April 2012. The growth plane of the BT films has been characterized by XRD patterns along the surface normal direction. Fig 1(b) clearly shows that the BT (0 0 0 1) plane is parallel to the Si (1 1 1) surface where Miller indices based on the hexagonal unit cell were used for the BT film and the cubic unit cell for the Si substrate. This growth orientation is consistent with previous reports.[8,12-16] Every peak position except around L = 4 and 13 satisfies the space group R-3m of BT (the reflections with H = K = 0 are allowed when only L = 3n, with n = integer) and the relative peak intensities are consistent with tabulated powder diffraction data. The additional peaks around L = 4 and 13 with the rocking width of several degrees in a specular condition appear in all four samples with their positions almost the same but with different intensities. One more additional peak around L = 25 has been observed in sample A1 and A2. By comparison with compounds consisting of Bi, Te, O, or Si,[17] the peak positions of L = 4 and 13 correspond to (1 1 0) and (4 2 0) reflections of $Te_{16}Si_{38}$[18] and L = 13 and 25 correspond to (0 0 4) and (-2 4 4) reflections of $Bi_2Te_2O_8$,[19] respectively. In addition, a strikingly strong enhancement of the forbidden Si (2 2 2) reflection is observed for all samples. We attribute the anomaly at L = 25 and possibly at L = 13 to oxidation of BT at the surface, the anomaly at L = 4 and possibly at L = 13 to the formation of $Te_{16}Si_{38}$ at the interface between BT and Si, and the forbidden Si (2 2 2) reflection to the large density of Te impurities in the Si substrate occupying mainly substitutional sites, which arises due to the large diffusion coefficient for Te in Si.[20] The reduced point group symmetry associated with Te impurities converts the Si (2 2 2) reflection from being forbidden to allowed. The measured lattice constants of all the Si substrates are 5.42 Å,[21] slightly smaller than that of undoped Si (5.43 Å), consistent with Te dilution of the Si substrate.[22] This finding is very important since it may explain the failure so far to observe metallic surface states in electrical transport measurements in BT films, which may have been obscured by shorts through the heavily Te-doped *n* type silicon substrates[23] and *n* type doped BT by Te vacancies and oxygen.[24]

The asymmetric (0 1 -1 L) reflections shown in Fig. 1(c) are consistent with the crystal structure of bulk BT. The lattice constants of the BT films obtained from these reflections are the same as the bulk values, a = b = 4.38 Å and c = 30.4 Å. The full width at half maxima (FWHM) of the rocking curves of less than 0.1° presented in Fig. 1(d) reveal excellent crystal quality of the BT films. Moreover these FWHM are almost independent of the reflection indicating, together with the specular reflection data, that the in-plane orientational order is as good as that associated with the stacking direction for BT. Fig 1(e) shows azimuthal scans of Si {3 3 1} and BT {0 1 -1 20} reflections in which the azimuthal angles are defined with respect to the [1 1 -2] and [0 1 -1 0] directions in the surface of Si and BT, respectively. The relation between the azimuthal peak positions of the Si substrate and the BT film demonstrates in-plane epitaxy of Si [1 1 -2] || BT [0 1 -1 0]. In these scans the peak positions of equivalent reflections are slightly different from those calculated using the crystal orientation matrix obtained by the surface normal and off-axis reflections due to a misalignment of the sample on the goniometer. To avoid missing other grains with different in-plane crystal directions due to this misalignment effect, we integrated the intensities of rocking curves within a range of 0.8° and confirmed that the entire film has a single orientation as shown in the inset of Fig. 1(e). The orientational match between Si and BT suggests that there is no amorphous layer of Te between Si and BT during the growth, whereby epitaxy would be lost. This is consistent with TEM studies of $Bi_2Se_3$ films grown on Si (1 1 1) that suggest the absence of an amorphous layer between the film and substrate during growth under optimum growth conditions.[25] The formation of an amorphous $SiO_2$ layer at the interface between $Bi_2Se_3$ and Si in uncapped films after the film is stored in ambient condition is attributed to oxygen diffusion through the film. Since the epitaxy between the film and substrate is locked during the growth in the absence of an amorphous interface, the formation of an amorphous layer at the interface afterwards does not alter the epitaxy between the film and the substrate. Our XRD measurements are not able to detect this amorphous layer since it tracks only crystal structures.

Fig. 1(f) presents the crystal orientations obtained from the XRD patterns. Although we use a 7×7 reconstructed Si (1 1 1) surface as a substrate to grow our BT films, the reconstruction is immediately removed and the surface becomes a Si 1×1 within seconds once we initiate the growth by exposing the substrate to Te flux.[26] When the lattice mismatch is large as in the case between Si and BT, the growth with supercell matching becomes favorable.[27] A Si supercell with 2 interatomic distances along the [1 -1 0] and [1 1 -2] directions is close to matching a BT supercell with 1 and 3 interatomic distances along the [2 1 -3 0] and [0 1 -1 0] directions, respectively. The lattice mismatches between the supercells are 0.9 % and 1.1 % along Si [1 -1 0] and [1 1 -2], respectively. The fact that these mismatches are small enables us to grow good epitaxial films.

The XRD data measured on the other samples show the same epitaxial relationships but the crystalline qualities are different. Fig. 2 shows the XRD patterns for the four samples. The reciprocal space maps presented in Fig. 2(a) show sharp spots corresponding to Si (1 1 1) and BT (0 0 0 3n) reflections at the same H and K and an absence of non-specular signals, indicating that the films have a single phase in common. However the widths of the spots depend on the sample. Not surprisingly the width of the rocking curve of BT films is similar to that of their Si substrates, as shown in Fig. 2(b). This similarity again shows the strong coupling between the film and its substrate.

Fig. 2(c) shows the x-ray diffraction patterns in the specular condition around BT (0 0 0 3). We can clearly see the Laue oscillations originating from the finite crystalline thicknesses of the films. The crystalline film thicknesses obtained from the oscillations around the Bragg peaks are 43.2±3.6 nm, 144.3±7.9 nm, 48.4±2.2 nm, and 123.3±8.8 nm for sample A1, A2, B1, B2, respectively. The crystallite sizes of the films can be also obtained by the Scherrer equation. Analyzing the BT (0 0 0 3) reflections using a Scherrer constant of 0.885[28] provides the sizes of 41.2±0.3 nm, 110.0±0.5 nm, 44.6±0.3 nm, and 113.9±0.3 nm, for sample A1, A2, B1, B2, respectively, where instrumental broadening was not corrected for. Within experimental error, the two methods are consistent with each other,[28] except for sample A2. Sample A2 happens also to show less clear oscillations. The Scherrer method is sensitive to the strain, mosaicity, and roughness of a film while the oscillation period of the thickness fringe is almost independent of these factors, although of course the visibility of the oscillations is strongly dependent on the roughness of the crystal boundaries. Therefore we conclude that the interface or surface of sample A2 is more distorted as well as much rougher and this lower film quality appears to be related to the especially low crystal quality of the Si substrate reflected in the broad rocking curve.

The XRD pattern on sample A1 has been compared with data measured twenty months earlier to check for ageing effects. The comparison shows three distinct differences while their overall profiles are the same, as shown in Fig. 3(a). First, the thickness fringes of the earlier data are clearer, which indicates that the crystalline boundaries of the film become rougher with time. Secondly, the width of the BT (0 0 0 15) reflection becomes a little wider with time so that it is found that the crystallite thickness of the film has been slightly reduced. The film thickness obtained from the fringes of the earlier data, which is 47.6±2.0 nm, is also larger than the later value of 43.2±3.6 nm. To study the roughness change over time Fourier transformation (FT) of the fringes around the BT (0 0 0 15) peak has been performed. Before carrying out the FT, the XRD intensities were normalized by a non-oscillatory profile for the (0 0 0 15) peak to isolate finite thickness from crystal structure effects. Fig. 3(b) clearly shows that the crystalline film became thinner and rougher. The thicknesses obtained from the peak positions are 47.4±0.3 nm and 46.4±0.7 nm, which agree to the thicknesses obtained by other methods within the errors, and the roughness estimated from the FWHMs are 4.6±0.1 nm and 8.4±0.2 nm for the measurements in August 2010 and April 2012, respectively. Finally, the absence of the extra peak at L = 13 in the earlier data, which is true for sample A2 as well, reveals that the formation of additional crystalline phases occurs after the growth and not during the growth.

Routine in-situ STM measurements after growth show atomically flat surfaces of our BT films, whereas Atomic Force Microscopy (AFM) images taken after ageing reveal islands with widths of 300-600 nm and heights of 30-80 nm on the surface of the samples. The XRD results suggest two materials for the composition of these islands, $Bi_2Te_2O_8$ and $Te_{16}Si_{38}$, but we have excluded $Te_{16}Si_{38}$ as it is unlikely that Si atoms move from the substrate to the surface through the BT film at room temperature. The crystalline size obtained from the peak widths of L = 13 and 25 are 30-50 nm, similar to the heights of the islands obtained from the AFM images. Therefore, our results suggest that oxidation of BT has occurred resulting in formation of crystalline $Bi_2Te_2O_8$ on the surface. Previous studies of the oxidation of BT surfaces using x-ray photoelectron spectroscopy (XPS) speculated that the oxide layer at the surface is amorphous,[29] consistent with no extra peak as in the XRD measurements obtained shortly after film growth. The later appearance of the extra peaks in XRD as well as tall structures in the atomic force micrographs could then be due to crystallization of some of the amorphous bismuth and tellurium oxides to $Bi_2Te_2O_8$ with time.

In addition to $Bi_2Te_2O_8$, we believe that crystalline $Te_{16}Si_{38}$ has formed at the interfaces with the Si substrates. The thickness of $Te_{16}Si_{38}$ obtained from the peak width of L = 4 is around 16 nm for sample B1 and B2 grown on B-doped Si substrates. For sample A1 and A2 grown on undoped Si, the thickness is not obtainable because the peaks at L = 4 are too weak. We speculate that the pre-existing dopants in the Si substrate might help to form the $Te_{16}Si_{38}$ phase, so that the samples on the p-doped Si exhibit much stronger signals than those on undoped Si. Besides the change in the XRD pattern from the film, we notice that the Si lattice constant measured in August 2010 is 5.43 Å for all samples, larger than the value of 5.42 Å measured in April 2012, which suggests that the Si substrates undergo a structural change after the film growth. We attribute the formation of $Te_{16}Si_{38}$ and structural change of Si to the volatility of Te and its diffusion into the Si substrate. The ageing effects, which consist of the reduction of the crystalline phase of BT, the roughening of the crystalline boundaries and film surface, formation of new crystalline phases at the film boundaries, as well as structural modification of the substrate can all be attributed to the tendency for BT to oxidize and the volatility of Te, and can have profound effects on the transport properties since it results in the creation of donors in BT via dissolved oxygen as well as Te vacancies and Te donors in Si, a topic of current interest because of applications to solar cells.[22]

In summary, we have characterized the crystal structure and epitaxy of BT films on Si (1 1 1) using x-ray diffraction. Our results show that the films are single crystal with an epitaxial relationship of BT [0 0 0 1] || Si [1 1 1] and BT [0 1 -1 0] || Si [1 1 -2]. We have found an unexpectedly good lattice match between BT and Si supercells resulting in a strong dependence of the crystalline quality of the films on those of their substrates. We observe that the film boundaries undergo ageing and that new crystalline phases form at the boundaries over time even though most of the film remains in the original single crystal phase. Finally, we found the crystal structure and composition of the Si substrates to be modified indicating that Te diffuses into the substrate. All of these effects will need to be considered when analyzing electrical data for films grown on silicon.

The work was supported by the UK EPSRC and Chinese National Science Foundation.

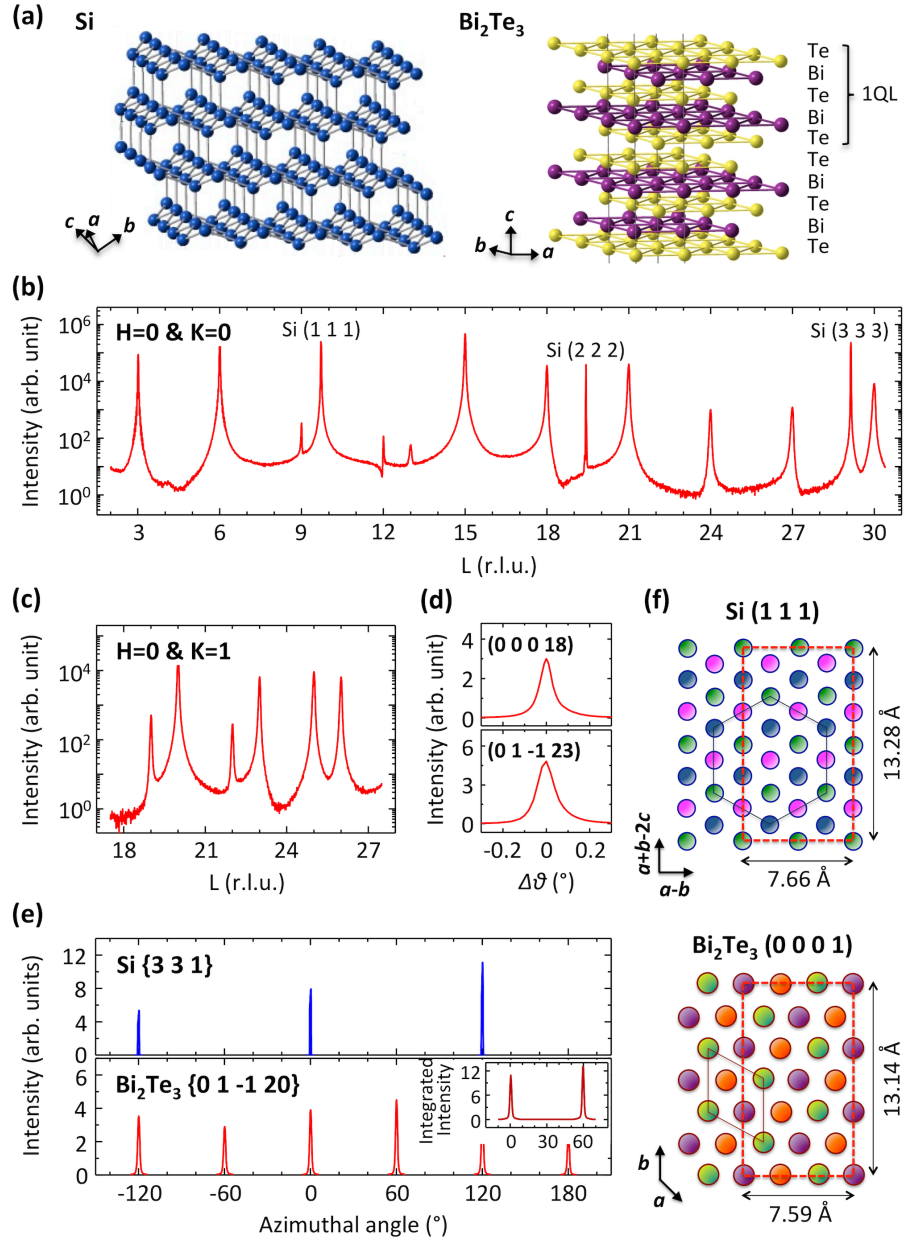

FIG 1. XRD measurements in April 2012 on sample B2. (a) The schematic crystal structures of Si and BT. (b) X-ray diffraction pattern along the surface normal direction, which is the [1 1 1] direction for Si substrates. Most of the peak positions and relative intensities are consistent with the space group R-3m of BT. A dip on the left side of BT (0 0 0 12) is due to interference with Si (1 1 1) reflection. The two extra peaks around L = 4 and L = 13 could not be indexed but their intensities are much smaller than those coming from BT. We observe a Si (2 2 2) reflection, which is a forbidden peak for pure silicon. (c) X-ray diffraction pattern measured as a function of L for the BT (0 1 -1 L) line of reflections. The forbidden reflections with L = 3n satisfy its space group. The lattice constants obtained from the peak positions are a = b = 4.38 Å and c = 30.4 Å consistent with the bulk values. (d) Rocking curves of the BT (0 0 0 18) and (0 1 -1 23) reflections. The FWHM of less than 0.1° shows excellent crystalline quality and is almost independent of the reflection index. (e) Azimuthal scans of Si {3 3 1} and BT {0 1 -1 20} reflections in which the azimuthal angle is defined with respect to the [1 1 -2] and [0 1 -1 0] directions on the surface of Si and BT, respectively. The relation between the azimuthal peak positions indicates an in-plane epitaxial relationship of Si [1 1 -2] ∥ BT [0 1 -1 0]. (Inset) Integrated intensities measured from rocking curves within theta range of 0.8°. Zero intensities except at 0° and 60° reveal that there are no other grains and the entire film has a single orientation. (f) The crystal structures of Si and BT projected to the surface normal planes. The different colored spheres represent atoms (Bi and Te) in different planes. A Si supercell with 2 interatomic distances along the [1 -1 0] and [1 1 -2] directions match with a BT supercell with 1 and 3 interatomic distances along the [2 1 -3 0] and [0 1 -1 0] directions, respectively, and the lattice mismatch is 0.9 % and 1.1 % along Si [1 -1 0] and [1 1 -2], respectively.

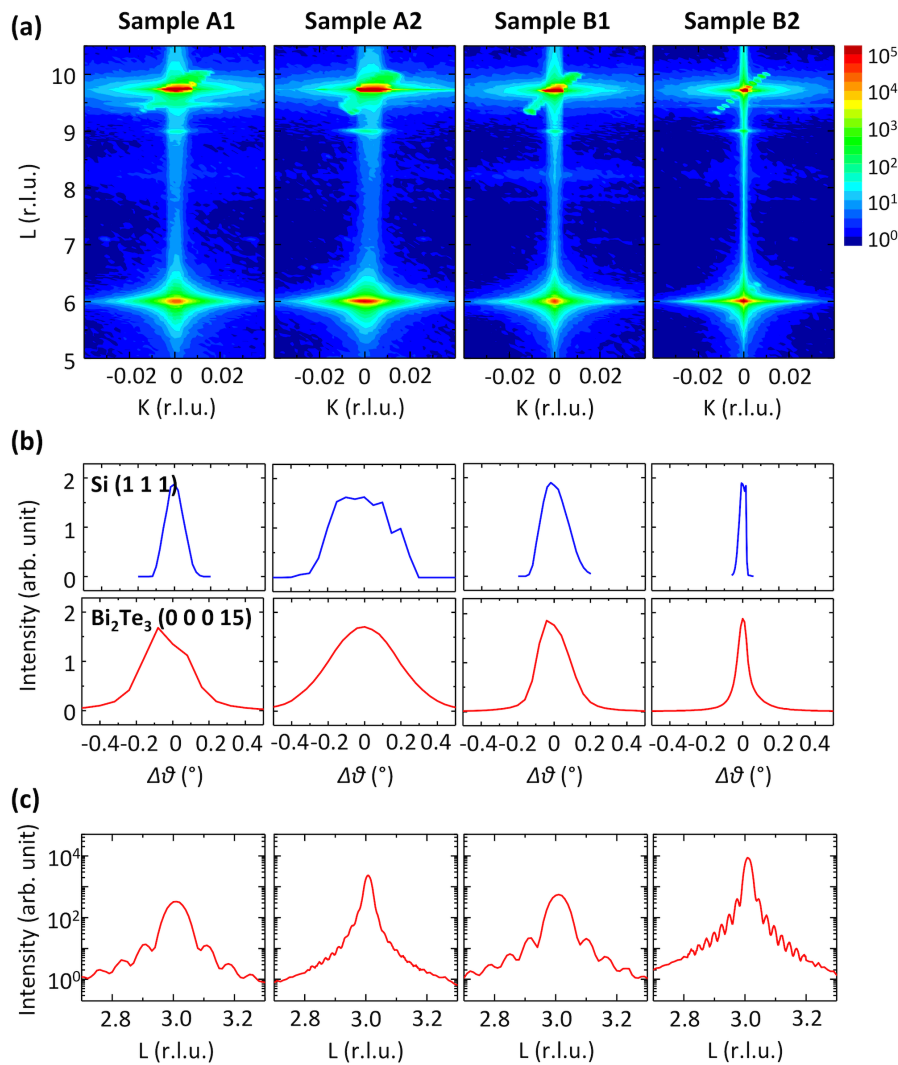

FIG 2. XRD measurements in April 2012 (a) Reciprocal space maps in the K-L plane with H = 0. Sharp spots corresponding to Si (1 1 1) and BT (0 0 0 3n) reflections at same H and K and the absence of non-specular signals reveal the BT films have a single phase in which the BT [0 0 0 1] direction is parallel to the Si [1 1 1] direction. The diagonal streaks across Si (1 1 1) spots originate from detector resolution. (b) Rocking curves of Si (1 1 1) and BT (0 0 0 15) reflections. Note the similarity of the FWHM of the film and Si substrate peaks. (c) X-ray diffraction patterns around BT (0 0 0 3) reflections. The film thicknesses obtained from the oscillation period are 43.2±3.6 nm, 144.3±7.9 nm, 48.4±2.2 nm, and 123.3±8.8 nm and the crystallite sizes of the films obtained by Scherrer equation (with Sherrer constant K = 0.885) are 41.2±0.3 nm, 110.0±0.5 nm, 44.6±0.3 nm, and 113.9±0.3 nm, respectively. The rougher and more distorted interface or surface of sample A2 than those of other samples could be related to the poor crystal quality of the Si substrate.

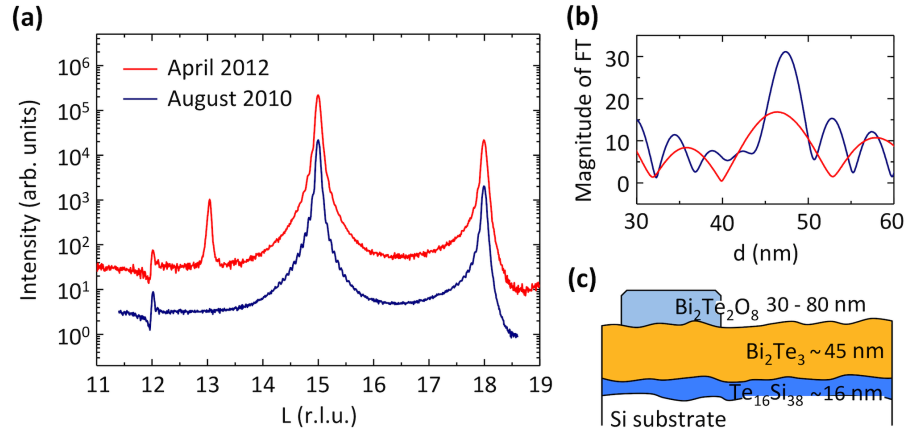

FIG. 3. (a) X-ray diffraction patterns with H = K = 0 for sample A1 measured in April 2012 and in August 2010. The curves have been shifted along the y-axis for clarity. There are two noticeable effects. There is a new forbidden peak at L=13 on April 2012, which was not present when the sample was measured in August 2010. Also, the oscillation amplitudes due to the finite film thickness around L=15 and L=18 are damped in April 2012 compared to August 2010. (B) Fourier transformation of the XRD fringes around the BT (0 0 0 15) reflections. Red and blue lines represent the results obtained from measurements in April 2012 and August 2010, respectively. The thicknesses obtained from the peak positions are 47.4±0.3 nm and 46.4±0.7 nm and the roughness estimated from the FWHMs are 4.6±0.1 nm and 8.4±0.2 nm for the earlier and later data, respectively. (c) Schematic diagram of the sample structure after the degradation. $Te_{16}Si_{38}$ has formed at the interface between the BT film and Si substrate and $Bi_2Te_2O_8$ islands have grown on the surface of BT. The numbers represent the layer thicknesses for sample B1.